\documentclass[iop]{emulateapj}

\usepackage{natbib}

\newcommand{\cs}{\langle\sigma_A\upsilon\rangle}

\shorttitle{Evidence for dark matter from galaxy clusters}
\shortauthors{Hektor, Raidal \& Tempel}

\begin{document}

\title{An evidence for indirect detection of dark matter\\ from galaxy clusters in Fermi $\gamma$-ray data}

\author{A. Hektor\altaffilmark{1}, M. Raidal\altaffilmark{2,3} and E. Tempel\altaffilmark{4}}
\affil{National Institute of Chemical Physics and Biophysics, R\"avala 10, 10143 Tallinn, Estonia}
\email{andi.hektor@cern.ch, martti.raidal@cern.ch, elmo@aai.ee}

\altaffiltext{1}{Helsinki Institute of Physics, P.O. Box 64, FI-00014, Helsinki, Finland}
\altaffiltext{2}{CERN, Theory Division, CH-1211 Geneva 23, Switzerland}
\altaffiltext{3}{Institute of Physics, University of Tartu, Estonia}
\altaffiltext{4}{Tartu Observatory, Observatooriumi 1, T\~oravere 61602, Estonia} 

\begin{abstract}
We search for spectral features in  Fermi-LAT gamma-rays coming from regions corresponding to eighteen brightest nearby galaxy clusters determined by the magnitude of their signal line-of-site integrals. We observe a double peak-like excess over the diffuse power-law background at photon energies 110~GeV and 130~GeV with the global statistical significance up to $3.6\sigma$, confirming independently earlier claims of the same excess from Galactic centre. Interpreting this result as a signal of dark matter annihilations to two monochromatic photon channels in galaxy cluster haloes, and fixing the annihilation cross section from the Galactic centre data,  we determine the annihilation boost factor due to dark matter subhaloes from data. Our results contribute to discrimination of  the dark matter annihilations from astrophysical processes and from systematic detector effects as the possible explanations to the Fermi-LAT excess.

\end{abstract}



\section{Introduction}

It is a prediction of the concordance cold-dark-matter cosmological model that galaxies and galaxy clusters are surrounded by a massive dark matter (DM) haloes. Firm evidence for the DM existence is coming from various gravitational effects in astrophysics and cosmology \citep{Bertone:2004pz}.
If the existing cosmological DM \citep{Komatsu:2010fb} is a thermal relic consisting of weakly interacting massive particles, DM annihilations into the standard model particles should provide indirect evidence of DM in cosmic ray experiments \citep{Cirelli:2010xx}.
Unfortunately the direct \citep{Aprile:2011hi} and indirect \citep{Cirelli:2010xx} searches for DM particles have all given either negative or contradictory results.
A notable exception to this result is the recent evidence for $\gamma$-ray excess with energy 
130~GeV \citep{Bringmann:2012vr,Weniger:2012tx,Tempel:2012ey,Su:2012ft} in the Fermi Large Area Telescope (LAT) \citep{Atwood:2009ez} data.
This excess originates predominately from a small region in the Galactic centre \citep{Tempel:2012ey,Su:2012ft} and may have a double peak structure \citep{Su:2012ft}.
Its global statistical significance  is between $4.5\sigma$ and $6.5\sigma$ \citep{Su:2012ft}, depending whether one or two peaks are fitted, and it is consistent with the Fermi-LAT bound on monochromatic photon lines from diffuse $\gamma$-ray data \citep{Ackermann:2012qk}.
Although Galactic background effects make statistical fluctuation in the $\gamma$-ray spectrum more likely than naively expected \citep{Boyarsky:2012ca}, the presence of a double peak in the background is very unlikely.
Although an option that the 130~GeV $\gamma$-ray excess is a fake detector effect is not entirely excluded, recent studies disfavour this possibility \citep{Hektor:2012ev,Finkbeiner:2012ez}.
It is likely that Fermi-LAT has either observed an astrophysical process that unexpectedly gives photon peak(s) 
or detected the DM annihilations into monochromatic photons.

To verify that the DM has been discovered indirectly, the $\gamma$-ray excess must either be confirmed by other experiments such as the planned  high-resolution experiments \mbox{CALET} or \mbox{TANSUO}, or to observe the 
same excess with Fermi-LAT from other known DM dominated objects.
The expected signal from nearby dwarf galaxies turned out to be too weak to check the 130~GeV $\gamma$-ray excess with Fermi-LAT \citep{GeringerSameth:2012sr}.
However, the galaxy clusters, the biggest nearby cosmological structures dominated by DM, are expected to be
much better objects for that purpose \citep{Huang:2011xr} because the DM annihilation signal from there should be amplified by a boost factor due to the existence of many DM subhaloes \citep{Springel:2008by,Springel:2008cc,Pinzke:2009cp,Anderson:2010df,Pinzke:2011,Gao:2011rf}.
There are large uncertainties in theoretical predictions of the boost factors \citep{Pieri:2007ir,Kuhlen:2008aw,Kamionkowski:2010mi}, numerical estimates vary from 10 to 10000. Experimental measurements are needed to discriminate between the different subhalo models.

Here we report on searches for spectral features in  Fermi-LAT $\gamma$-rays from regions corresponding to nearby galaxy clusters determined by the magnitude of
their signal line-of-sight integrals ($J$-factors). We observe a double peak-like excess at photon energies 110~GeV and 130~GeV over the diffuse power-law background with statistical significance up to $3.6\sigma$, confirming independently the earlier claims of excess from the Galactic centre. 
Interpreting this result as a signal of DM annihilations into two channels with monochromatic final-state photons, and fixing the annihilation cross section from Galactic centre data, we determine the annihilation boost factor due to galaxy cluster subhaloes.

\section{Data}

\begin{figure}
\includegraphics[width=0.5\textwidth]{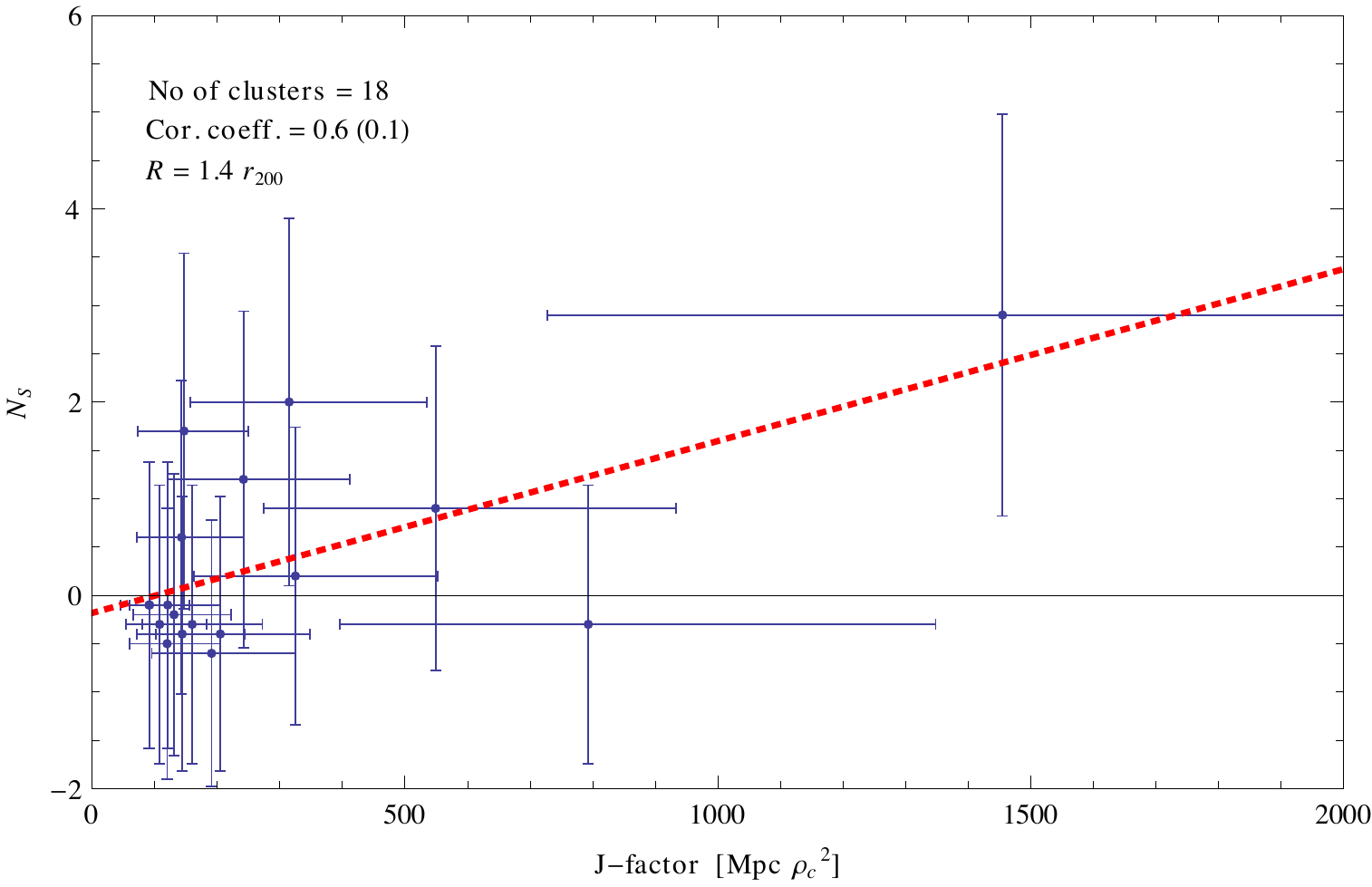}
\caption{Correlation between the estimated number of signal photons and the J-factors of galaxy clusters.}
\label{fig:jfactor}
\end{figure}

In this work we search for spectral features in the $\gamma$-ray spectrum originating from the known galaxy clusters in Fermi-LAT data. For that, we work with a set of galaxy clusters which parameters are reliably known in the literature. We compute their $J$-factors according to \citet{Pinzke:2011} that allows us to select clusters according to their expected contribution to the DM annihilation signal. Coordinates, masses, distances and radii for the 18 galaxy clusters with largest $J$-factors are collected in Table~\ref{tab1}.
Because of limited statistics we cannot study signal from every cluster individually. Instead we sum up all the photons coming from the directions of these clusters and study the spectrum of stacked flux.
Current determinations of the galaxy cluster DM halo parameters suffer from large uncertainties. Despite of that we do observe a correlation between the $J$-factors of the galaxy clusters and the number of signal photons from them (see Fig.~\ref{fig:jfactor}), supporting our claim that the signal originates from galaxy clusters.
We reanalyse the Galactic centre excess with new LAT resolution and find the double peak in the same position, supporting our claim that the two spectra both signal DM annihilations. Assuming this, we determine the DM annihilation boost factor in galaxy clusters from Fermi LAT data.

In the present analysis we consider the public Fermi-LAT photon event data of 218 weeks within energy region from 20 to 300~GeV. We apply the quality-filter cuts, as recommended by the Fermi-LAT team, as we did in our previous study \citep{Tempel:2012ey}. We make use of the \mbox{ULTRACLEAN} events selection (Pass~7 Version~6), in order to minimise potential systematical errors. The selection of events as well as the calculation of exposure maps is performed using the Fermi \mbox{ScienceTools}. 
The most important improvement compared to our previous work \citep{Tempel:2012ey} is the usage of new improved Fermi-LAT energy resolution \citep{Ackermann::2012kca}.

To avoid effects of point sources we exclude photons that are within an energy-independent cut radius of each source. We used all 1873 sources from the LAT 2-year point source catalog \citep{Abdo:2010ru}. The cut radius is chosen to be 0.2$^\circ$ \citep{Ackermann:2012qk}. We also tested the radii 0$^\circ$, 0.15$^\circ$, 0.25$^\circ$ and 0.5$^\circ$ and find no effect on the final result.

\section{Methods}

To estimate the photon spectrum related to the clusters we select photons that are within an energy independent radii around the centre of each cluster. We will denote it as the region of interest (ROI) below. The annihilation signal should arise predominantly within  $r_{200} (\approx r_{\rm vir})$  in each cluster. The boosting effect due to halo substructures  should make the signal spatially flat \citep{Pinzke:2009cp}.  
The morphology of expected signal from a single main halo without substructure should be very different: it should arise from a small central region and must have very cuspy nature \citep{Ackermann:2010rg}. However, in order to determine the stability signal significance over background, we need to consider larger and different regions than $r_{200}$ around clusters.

We studied both the radii $r_{200}$ dependent and independent ROIs around the clusters. We found that the result is independent of how the background photons are included -- signal peaks are not affected while the background fluctuates within estimated errors. Thus the simplest choice of ROI is the equal radii for all clusters. We considered radii $R=$ 3, 4, 5, 6 and 8 degrees for ROI and found that starting from smaller values of $R$, at $5^\circ$ the significance reached maximal value and remained stable for larger radii. We also found that galaxy clusters with small $J$-factors mostly contribute to the background reducing the signal background ratio. Starting from smaller number of clusters, the significance reached the maximum at 5$\dots$7 clusters and remained stable beyond that. For numerical results we therefore consider only the 18 most relevant galaxy clusters presented in Table~\ref{tab1}.

To compute  the $\gamma$-spectrum we sum up all photons from the selected cluster regions. The spectrum is calculated by the kernel smoothing as described in detail by \citet{Tempel:2012ey}. The characteristic kernel size is chosen based on the Fermi-LAT energy respond function. We calculated the spectrum in logarithmic and linear energy scales and used different kernel functions and sizes. The results are rather insensitive to the exact kernel function and size, showing that this kernel smoothing method is rather robust.

To estimate the signal significance we select $N$ random cluster-size regions in sky and find the $\gamma$-spectrum for them. To avoid the crowded region at the Galactic plane and centre of the Galaxy we exclude the region $|b|<5^\circ$ from the study \citep{Ackermann:2012qk}. It means the border of a randomly selected region can not be closer to the Galactic plane than $|b|=5$. We tested different sizes of the excluded regions: $|b|<$ 5, 10 and 15 degrees having neglectable effect on results.
For significance estimation of the spectral features we repeat the procedure 100,000 times for all the selected cases of radii to get the distribution of spectra. Based on the 100,000 Monte Carlo realisations we estimated the confidence limits of the spectra.

To estimate the boost factor, we use the relation between the number of signal photons $N_{\rm signal}$ within solid angle $\Delta \Omega$ and the properties of DM particle:
\begin{equation}
 \frac{N_{\rm signal}}{T_{\rm exp}} = \frac{1}{4\pi} \frac{\cs}{2 m_{\rm DM}^2} \, B \, J_{\Delta\Omega} \, N_{\rm prod},
\end{equation}
where $T_{\rm exp}$ is the exposure  of cluster region $\Delta \Omega$, $\cs$ is the averaged cross-section, $m_{\rm DM}$ is the mass of DM particle and $N_{\rm prod}$ is the number of produced photons per annihilation (in case of non-self-conjugated particle there is addition factor two in front of $m_{\rm DM}$). The $J$-factor $J_{\Delta\Omega}$ is defined by the line-of-sight integral
\begin{equation}
 J_{\Delta\Omega} = \int_{\Delta\Omega} d\Omega \, J(\Omega) = \int_{\Delta\Omega} d\Omega \, \int_{\text{l.o.s}} ds \, \rho^2(s,\Omega),
 \label{J}
\end{equation}
where $\rho(s,\Omega)$ is the density profile of DM in galaxy cluster. The parameters of the main DM halo is considered from \citet{Ackermann:2010rg}. We take the annihilation cross-section to photons to be $0.1 \times \cs_{\rm th}$, where $\cs_{\rm th} = 3 \times 10^{-26}$ cm$^3$ s$^{-1}$ is the standard thermal cross-section, as computed from the $\gamma$-ray line signal from the Galactic centre \citep{Weniger:2012tx, Tempel:2012ey}.

\section{Results}

\begin{figure*}
\includegraphics[width=1.0\textwidth]{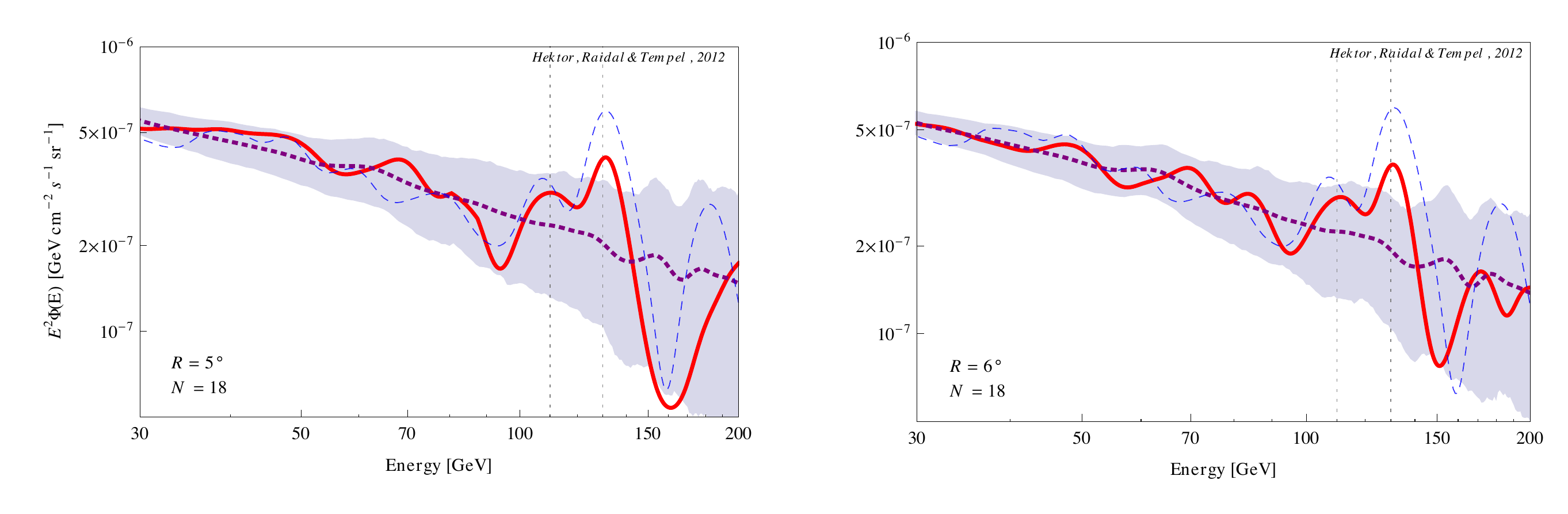}
\caption{Measured $\gamma$-ray spectra for fixed $R=5^\circ,\, 6^\circ$ 
regions around the 18 galaxy clusters as functions of photon energy (red solid curve).  The purple dashed line shows a fit to the background
together with its 95\% CL error band. The blue dashed curve  shows the reduced signal from Galactic centre for comparison.}
\label{fig1}
\end{figure*}

Fig.~\ref{fig1} shows the measured spectrum of $N=18$ galaxy clusters for two choices of fixed radii, $5^\circ,$ and $ 6^\circ,$ together with the 95\% CL background estimated with Monte Carlo (grey band). Reduced signal from the Galactic centre is also presented for comparison. Due to the improved energy resolution (smaller smoothing kernel) we observe double peak structure in both spectra. The first peak is at 110~GeV and the second at 130~GeV, consistently with two DM annihilation channels to $\gamma Z$ and $\gamma\gamma,$ respectively. This is consistent with a generic prediction of gauge theories. Looking for the double-peak excess with at least the same strenght as observed lines (110~GeV and~130 GeV) in the whole MC sample and energy region (20\dots300~GeV) implies the global significance of the double-peak signal with $3.6\sigma$.

Table~\ref{tab2} presents the number of photons in the signal region divided into 5~GeV energy bins for different angular regions around the galaxy clusters. We see that the photons are clustered around two energies (110 and 130~GeV) corresponding to the two peaks in Fig.~\ref{fig1}. For larger values of considered radii the statistics increases, giving an optimal signal-over-background ratio for $5^\circ$ regions. For larger radii, such as $8^\circ$, we observe slight reduction of the signal significance, probably due to a fact the background starts dominating and it is not stable for so large ROI.

Assuming that the signal originates from DM annihilation, we estimate the boost factor due to DM substructures in the galaxy cluster. If this assumption is correct, this is the first real measurement of the boost factor from experimental data.

Considering the radius $R=5^\circ$, the boost factor of substructure turns out to be $790 \dots 8200$ at 68\% CL, where the variation is estimated with bootstrap analysis. The error are very large and we note that this error is only statistical. In addition, other sources of error, the distance and mass of clusters, averaging over the selected clusters, the main DM profile of the clusters etc., are not taken into account.  In addition, the measured value of DM annihilation cross section to the photon channels in the Galactic centre, $0.1 \times \cs_{\rm th}$, depends on uncertainties of the Galactic DM halo and it has accuracy in order of magnitude. Taking into account all those uncertainties, the present experimental precision does not allow to determine the boost factor numerically with a meaningful statistical significance. In order to distinguish between different DM subhalo models, new more precise measurements are needed. We stress that the presence of double-peak feature in Fermi LAT data are not affected by this uncertainties.

The galaxy clusters, as well as the Galactic centre, are fixed objects dominated by DM, thus our result does not suffer from statistical fluctuations related to scanning and choosing arbitrary regions of the sky, nor from possible astrophysical effects from the Galactic disc. The fact that the two signals from unrelated regions of sky, from the Galactic centre and from the locations of galaxy clusters, give a double peak-like excesses that precisely coincide (see Fig.~\ref{fig1}) suggests that this is not a statistical fluctuation. This result disfavours astrophysical explanation to the excess since astrophysical processes should not be exactly the same in galaxy clusters and in the Galactic centre. Our result implies that most plausibly the DM of Universe has been discovered via indirect detection by Fermi-LAT.

\section{Conclusions}

We have found that Fermi-LAT data shows a double peak-like excess of $\gamma$-rays with energies 110~GeV and 130~GeV from the nearby galaxy clusters. The maximal global significance of the excess is  $3.6\sigma$. Our result provides an {\it independent} confirmation of the previously claimed   $\gamma$-ray excess  from the Galactic centre and supports the interpretation of the excess due to DM annihilations into two monochromatic photon channels. Making this assumption, and fixing the DM annihilation cross section from the Galactic centre data, we found the boost factor for DM annihilations due to DM substructures in the galaxy cluster from Fermi-LAT data. This is the first measurement of DM boost factor from real data and has potentially important implications for understanding the DM substructures in haloes. Unfortunately the related uncertainties at present are quite large. More data is needed to discriminate between different theoretical models of DM substructure as well as to discriminate the DM interpretation of the excess from the possible astrophysical origin and systematic detector effects.

The 130~GeV DM is kinematically accessible in the LHC experiments and should be searched for.

\acknowledgments

We thank L.~Bergstrom, J.~Conrad, D.~Finkbeiner and C.~Weniger for numerous communications. This work was supported by the ESF grants 8090, 8499, 8943, MTT8, MTT59, MTT60, MJD52, MJD272, by the recurrent financing projects SF0690030s09, SF0060067s08 and by the European Union through the European Regional Development Fund.

\bibliographystyle{apj}


\newpage

\begin{table*}[t]
\caption{The most relevant galaxy clusters according to their $J$-factors.}
\begin{center}
\begin{tabular}{lcccccc}
\hline\hline
{Cluster} & {$l$ (deg)} & {$b$ (deg)} & {$M_{\rm 200}$ ($10^{14}M_\odot$)} & {$D$ (Mpc)} & $r_{\rm 200}$ (deg) & $J$ (Mpc $\rho_c^2$) \\[0.5ex]
\hline
Virgo & -76.2 & 74.5 & 6.9 & 17.2 & 5.6 & 1465\\
Fornax & -123.3 & -53.6 & 2.4 & 19.77 & 3.7 & 793\\
M49 & -73.1 & 70.2 & 1.4 & 18.91 & 3.24 & 549\\
NGC4636 & -62.3 & 65.5 & 0.5 & 15.89 & 2.74 & 325\\
A3526 (Centaurus) & -57.6 & 21.6 & 5.3 & 44.46 & 2.15 & 315\\
Ophiuchus & 0.6 & 9.3 & 40.5 & 122.51 & 1.53 & 242\\
A1060 (Hydra) & -90.4 & 26.5 & 4.1 & 49.25 & 1.78 & 205\\
NGC5813 & -0.8 & 49.8 & 1. & 27.55 & 1.97 & 191\\
A3627 (Norma) & -34.7 & -7.3 & 7.2 & 70.69 & 1.49 & 160\\
Perseus & 150.4 & -13.4 & 8.6 & 79.48 & 1.41 & 147\\
AWM7 & 146.3 & -15.6 & 7.2 & 74.64 & 1.41 & 144\\
ANTLIA & -87.1 & 19.2 & 2.8 & 50.13 & 1.54 & 143\\
Coma & 58.1 & 88. & 12.9 & 101.14 & 1.27 & 131\\
A1367  & -125.2 & 73. & 10.1 & 94.05 & 1.25 & 121\\
NGC5846  & 0.4 & 48.8 & 0.5 & 26.25 & 1.66 & 120\\
NGC5044  & -48.8 & 46.1 & 1.1 & 38.81 & 1.46 & 108\\
A2877  & -66.9 & -70.9 & 9.5 & 105.13 & 1.1 & 92\\
3C129 & 160.4 & 0.1 & 7.8 & 97.15 & 1.11 & 91\\[1.0ex]
\hline
\end{tabular}
\end{center}
\label{tab1}
\begin{list}
    {}{} 
    \item[] \textbf{Notes.}
	The Galactic coordinates ($l,b$) are taken from the NASA/IPAC Extragalactic Database http://nedwww.ipac.caltech.edu/ and other data is collected from~\cite{Chen:2007sz,Han:2012uw,Reiprich:2001zv,Pinzke:2011}.
	 The 2$\sigma$ errors for $J$-factor are estimated to be $\pm^{60\%}_{40\%}$.
\end{list}
\end{table*}

\begin{table*}[t]
\caption{Numbers of $\gamma$-ray events binned in 5 GeV intervals for different angular regions around the selected clusters.}
\begin{center}
\begin{tabular}{lcccccccccc}
\hline\hline
$E$ ($\pm$2.5 GeV) & \scriptsize{102.5} & \scriptsize{107.5} & \scriptsize{112.5} & \scriptsize{117.5} & \scriptsize{122.5} & \scriptsize{127.5} & \scriptsize{132.5} & \scriptsize{137.5} & \scriptsize{142.5} & \scriptsize{147.5}\\[0.5ex]
\hline
$N$ ($R=3^\circ$) & 2 & 1 & 0 & 2 & 1 & 1 & 4 & 1 & 1 & 1\\
$N$ ($R=4^\circ$) & 2 & 1 & 4 & 3 & 2 & 4 & 6 & 1 & 1 & 1\\
$N$ ($R=5^\circ$) & 5 & 5 & 10 & 4 & 6 & 8 & 7 & 3 & 1 & 2\\
$N$ ($R=6^\circ$) & 8 & 7 & 15 & 6 & 7 & 10 & 9 & 5 & 1 & 2\\
$N$ ($R=8^\circ$) & 16 & 16 & 24 & 9 & 10 & 16 & 11 & 9 & 1 & 3\\[1.0ex]
\hline
\end{tabular}
\end{center}
\label{tab2}
\end{table*}

\end{document}